\documentclass[conference]{IEEEtran}
\IEEEoverridecommandlockouts

\usepackage{cite}
\usepackage{amsmath,amssymb,amsfonts}
\usepackage{algorithmic}
\usepackage{graphicx}
\usepackage{textcomp}
\usepackage{xcolor}
\usepackage{multirow}
\usepackage{pifont}
\usepackage{makecell}
\usepackage{subcaption}
\usepackage{threeparttable}
\usepackage{booktabs}
\usepackage{url}
\usepackage{hyperref}
\newcommand{\cmark}{\ding{51}}
\newcommand{\xmark}{\ding{55}}
\def\BibTeX{{\rm B\kern-.05em{\sc i\kern-.025em b}\kern-.08em
    T\kern-.1667em\lower.7ex\hbox{E}\kern-.125emX}}
\begin{document}

\title{Comparing Unsupervised and Supervised Semantic Speech Tokens: A Case Study of Child ASR
}


\author{
\IEEEauthorblockN{
Mohan Shi,
Natarajan Balaji Shankar, 
Kaiyuan Zhang,
Zilai Wang,
Abeer Alwan}

\IEEEauthorblockA{
\textit{Department of Electrical and Computer Engineering} \\
\textit{University of California Los Angeles}\\
Los Angeles, USA \\
\textit{\{shimohan, balaji1312, kaiyuanzhang, zilaiwang2001\}@ucla.edu, alwan@ee.ucla.edu}
}
}




\maketitle

\begin{abstract}
Discrete speech tokens have gained attention for their storage efficiency and integration with Large Language Models (LLMs). They are commonly categorized into acoustic and semantic tokens, with the latter being more advantageous for Automatic Speech Recognition (ASR). Traditionally, unsupervised K-means clustering has been used to extract semantic speech tokens from Speech Foundation Models (SFMs). Recently, supervised methods, such as finite scalar quantization (FSQ) trained with ASR loss, have emerged for speech generation. Both approaches leverage pre-trained SFMs, benefiting low-resource tasks such as child ASR.

This paper systematically compares supervised and unsupervised semantic speech tokens for child ASR. Results show that supervised methods not only outperform unsupervised ones but even unexpectedly surpass continuous representations, and they perform well even in ultra-low bitrate settings. These findings highlight the advantages of supervised semantic tokens and offer insights for improving discrete speech tokenization.
\end{abstract}

\vspace{0.2cm}
\section{Introduction}
\vspace{0.1cm}

In recent years, the rapid advancement of deep learning~\cite{X12a} has driven remarkable progress in speech processing, particularly in Automatic Speech Recognition (ASR)~\cite{GravesFGS06,abs-1211-3711,ChorowskiBSCB15,DongXX18,GulatiQCPZYHWZW20,KimWPPSHW22,li2022recent}. Conventional ASR systems typically rely on continuous acoustic features such as Mel-Frequency Cepstral Coefficients (MFCCs) or Filterbanks (Fbanks), or leverage high-dimensional representations learned via self-supervised or data-driven methods~\cite{BaevskiZMA20,HsuBTLSM21,ChenWCWLCLKYXWZ22,YangCCLLLLSCLHT21}.
More recently, the use of discrete speech tokens has attracted growing attention~\cite{ZeghidourLOST22,DefossezCSA23,ChangYFM023,ChangYCJLMSST0F24,YangSDM0P024,chang24b_interspeech}. These methods encode speech into sequences of discrete tokens, serving as compact and symbolic representations for downstream tasks. Discrete tokens offer several key advantages: (1) they significantly reduce storage and computational costs due to their low-bitrate nature; and (2) they enable seamless integration with large language models~\cite{abs-2301-02111,abs-2407-05407,abs-2412-10117}, thereby facilitating unified speech-language modeling in multi-modal frameworks.


In the literature, discrete speech tokens are generally classified into two main categories: \textit{acoustic} and \textit{semantic} tokens. Acoustic tokens are typically learned through signal reconstruction objectives, such as those used in neural audio codecs~\cite{ZeghidourLOST22,DefossezCSA23}. These tokens are effective in capturing low-level acoustic details but often lack higher-level linguistic abstraction, thus conveying limited semantic content.
In contrast, so-called 'semantic' tokens~\cite{ChangYFM023,YangSDM0P024,chang24b_interspeech} are extracted from the representations of Speech Foundation Models (SFMs)~\cite{BaevskiZMA20,HsuBTLSM21,ChenWCWLCLKYXWZ22}. As a result, semantic tokens inherently encode richer linguistic and contextual information. This makes them particularly well-suited for high-level speech tasks such as ASR, where capturing semantics is more crucial than preserving fine-grained acoustic fidelity.


Traditionally, unsupervised K-means clustering~\cite{Lloyd82} has been the predominant approach for extracting semantic tokens from the intermediate representations of SFMs~\cite{ChangYFM023,YangSDM0P024,chang24b_interspeech}. This method discretizes the continuous speech representations by assigning each frame-level embedding to one of the cluster centroids, which are learned from the training set. While simple and effective, it does not incorporate task-specific objectives and thus may overlook semantically important distinctions.
More recently, attention has been directed toward leveraging supervised learning objectives to train quantizers based on SFM representations~\cite{abs-2407-05407,abs-2412-10117}. {Among these methods, the Finite Scalar Quantization (FSQ) technique~\cite{MentzerMAT24}, which was proposed as a core component of the supervised semantic speech ($S^3$) tokenizer~\cite{abs-2412-10117}, demonstrates higher codebook utilization than traditional Vector Quantization (VQ)~\cite{abs-2407-05407}. The supervised FSQ tokenizer, optimized with an ASR loss during training, has demonstrated strong empirical performance in text-to-speech (TTS)~\cite{abs-2412-10117} as the learned tokens retain high-level semantic information beneficial for next-token prediction.}




\begin{figure*}[t]
    \centering
    \captionsetup[subfigure]{justification=centering} 
    \subfloat[Extraction pipeline of unsupervised semantic speech tokens using K-means clustering.]{
        \includegraphics[width=0.98\linewidth]{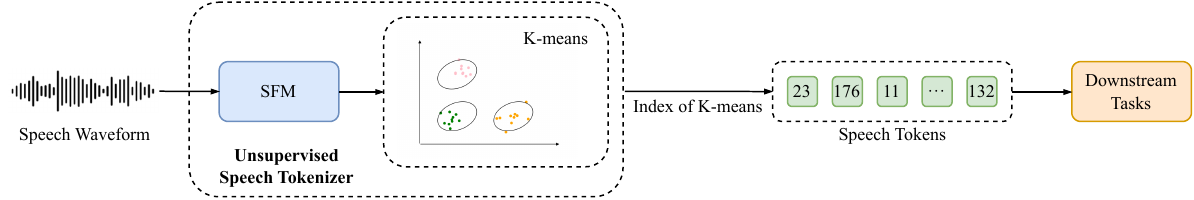}
        \label{subfig:unsupervised}
        \vspace{0.1cm}
    }    
    \vskip 1em
    
    \subfloat[Extraction pipeline of supervised semantic speech tokens using Finite Scalar Quantization (FSQ).]{
        \includegraphics[width=0.98\linewidth]{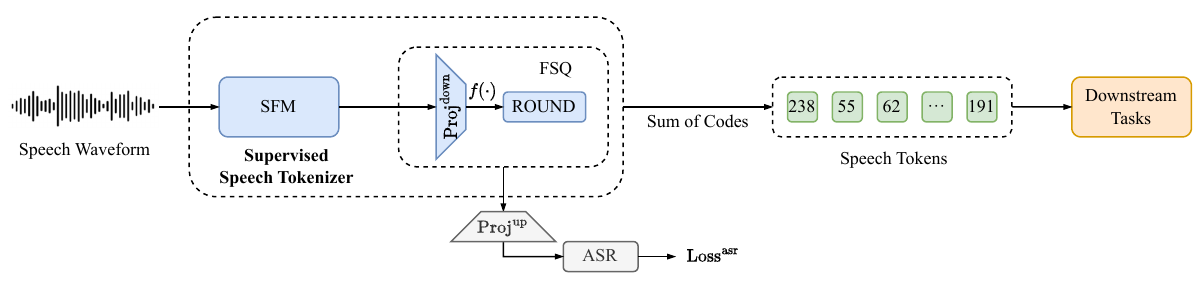}
        \label{subfig:supervised}
    }
    \vspace{0.5cm}
    \caption{Schematic illustration of the extraction pipeline for both unsupervised and supervised semantic speech tokens. $f(\cdot)$ denotes the bounding function.}
    \label{fig:tokenizer}
    \vspace{0.2cm}
\end{figure*}

Both unsupervised and supervised discrete speech tokens leverage the pre-trained knowledge embedded in SFMs, making them particularly suitable for low-resource ASR scenarios, such as children's speech recognition~\cite{WardCBBSVWZB11,ShobakiHC00}.
Prior work~\cite{fan24b_interspeech} has benchmarked children's ASR using SFMs, while other studies have explored unsupervised K-means-based semantic tokens for this domain~\cite{sukhadia24_interspeech, dutta2025exploring}. However, these unsupervised approaches often result in degraded ASR performance compared to continuous features.
Despite growing interest, there has been no systematic evaluation comparing discrete semantic tokens with continuous features for children's ASR. In particular, the relative strengths of unsupervised versus supervised semantic tokens for ASR remain underexplored. A comprehensive investigation is thus needed to better understand the trade-offs between token types and their potential for improving ASR.



To address these gaps, this study systematically compares semantic speech tokens derived from unsupervised K-means and supervised FSQ with continuous representations from SFMs, in the context of child ASR as a representative low-resource setting. 
Experimental results and in-depth analysis uncover unexpected performance and yield meaningful insights into speech tokenization.




In particular, the main contributions of this paper are:

\begin{itemize}
  \item We conduct a comprehensive comparison between unsupervised and supervised semantic speech tokens on the ASR task across multiple child speech datasets. {Our analysis covers not only ASR performance but also bitrate efficiency, codebook distribution, and cross-domain generalization, providing new insights into the characteristics of these tokenization methods.}
  \item Our results demonstrate that supervised FSQ tokens not only outperform unsupervised K-means tokens but also, unexpectedly, \textbf{surpass continuous high-dimensional representations extracted from in-domain SFMs}, revealing a previously underexplored advantage of supervised discrete tokenization.
  \item Through in-depth analysis of codebook usage and extended experiments, we demonstrate that \textbf{discrete semantic speech tokens retain competitive ASR performance even at ultra-low bitrates}, underscoring their potential for efficient and effective speech representation. {We further evaluate explore their generalizability across different speaker styles and age groups, revealing key limitations and future research directions.}
\end{itemize}

\section{Semantic Speech Tokens}
Semantic speech tokens, commonly referred to as such in the literature despite capturing a mix of phonetic and linguistic cues~\cite{choi24b_interspeech}, are derived from speech foundation models and are primarily used in semantics-oriented tasks such as ASR. Their extraction methods broadly fall into two categories: unsupervised and supervised approaches, as illustrated in Figure~\ref{fig:tokenizer}. For clarity and consistency, we adopt the term \textit{semantic} tokens throughout this paper to distinguish them from low-level acoustic tokens.

\subsection{Unsupervised Semantic Tokens Based on K-means}
K-means clustering~\cite{Lloyd82} is one of the most widely used unsupervised methods for extracting semantic speech tokens. The overall extraction pipeline is illustrated in Figure~\ref{subfig:unsupervised}. Given a speech training dataset \( X \), we first extract continuous representations using a pre-trained speech foundation model (SFM):
\begin{equation}
H = \text{SFM}(X)
\label{eq:sfm}
\end{equation}
where \( H = \{ h_n \in \mathbb{R}^d \mid n = 1, \dots, N \} \) represents the frame-wise continuous representations, with \( d \) denoting the feature dimension and \( N \) the total number of frames.

Next, we train a K-means model on \( H \):  
\begin{equation}
C = \text{K-means}(H)
\end{equation}
where \( C = \{ c_k \in \mathbb{R}^d \mid k = 1, \dots, K \} \) represents the \( K \) cluster centroids computed based on the Euclidean distance.  

For each frame, we assign the closest cluster index to obtain the discrete token set \( Z^\text{km} = \{ z_n^\text{km} \in \mathbb{Z} \mid n = 1, \dots, N \} \):  
\begin{equation}
z_n^\text{km} = \mathop{\text{argmin}}_{k \in \{1, \dots, K\}}  \lVert x_n - c_k \rVert^2
\end{equation}
The clustering model is trained on the training set and applied consistently across all subsets. The resulting tokens serve as discrete inputs for downstream tasks.

\vspace{0.2cm}
\subsection{Supervised Semantic Tokens Based on FSQ}
\vspace{0.1cm}
Unlike unsupervised tokenizers that rely on K-means clustering to obtain discrete tokens, the Supervised Semantic Speech ($S^3$) tokenizer~\cite{abs-2407-05407,abs-2412-10117} employs the ASR loss during training to guide the speech discretization process, as illustrated in Figure~\ref{subfig:supervised}.

Specifically, given a speech training dataset \( X \), the SFM first generates intermediate representations \( H \), as shown in Equation~\eqref{eq:sfm}. These representations are then passed through the finite scalar quantization (FSQ)~\cite{MentzerMAT24} module for discretization.

In the FSQ module, the intermediate representations \( H \) are first projected into a low-rank space of dimension \( d^\text{low} \). Each dimension is then quantized into a fixed number of discrete values using a bounding function \( f(\cdot) \) (e.g., \(\tanh\)) followed by a rounding operation.
\begin{align}
& H^\text{down} = \text{Proj}^\text{down}(H) \\
& H^\text{code} = \text{ROUND}(f(H^\text{down}))
\end{align}
The resulting quantized representation $H^\text{code} = \{ h_n^\text{code} \in \mathbb{Z}^{d^\text{low}} \mid n = 1, \dots, N \}$ forms an implicit codebook. Given the hyperparameters of FSQ, denoted as $\mathcal{L} = \left[L_0, L_1, \dots, L_{d^\text{low}-1}\right]$,  
where $d^\text{low}$ represents the number of low-rank channels, and $L_i$ corresponds to the number of quantization levels per channel. Each quantized code $h_{n,i}^\text{code}$ is an integer within the range $[0, L_i)$.  

The final speech token $z_n^\text{fsq}$ is computed as a weighted sum of the quantized low-rank codes $h_n^\text{code}$:
\begin{equation}
z_n^\text{fsq} = h_{n,0}^\text{code} + \sum_{i=1}^{d^\text{low}-1} (h_{n,i}^\text{code}\prod_{j=0}^{i-1}L_j)
\label{eq:fsq}
\end{equation}
The size of the implicit codebook is determined by the product of all elements in $\mathcal{L}$.
During tokenizer training, the quantized representation $H^\text{code}$ is projected back to its original dimensional space using an up-projection function:
\begin{equation}
\hat{H} = \text{Proj}^\text{up}(H^\text{code})
\end{equation}
\vspace{0.2cm}
The reconstructed representation, $\hat{H}$ is then fed into an ASR module to generate the ASR hypothesis, which is then compared with the ground truth transcriptions to compute the ASR loss for optimizing the supervised speech tokenizer:
\begin{align}
&\hat{Y} = \text{ASR}(\hat{H}) \\
&Loss^\text{fsq} = \text{Loss}^\text{asr}(\hat{Y},Y)
\end{align}
where $\hat{Y}$ and $Y$ denote the predicted and ground truth transcriptions, respectively. After training the entire system with the ASR loss, the discrete speech tokens $z_n^\text{fsq}$ are extracted using Equation~\eqref{eq:fsq} and can then be utilized for downstream tasks.

\begin{table*}[t]
\centering
\caption{WER (\%) comparison of various methods on MyST and OGI. The first three rows show results from previous works, where fine-tuned SSL SFMs were directly used for inference as reference baselines. The remaining rows present results obtained by feeding different input features (continuous or discrete) into the downstream ASR model. Bitrate (bits/sec) serves as a reference for input data efficiency. The downstream ASR adopts an E-Branchformer with a CTC-only architecture and greedy decoding. The bold font indicates the best performance.}
\resizebox{0.95\textwidth}{!}{
\begin{tabular}{c|c|c|cc|cc|c}
\toprule
    \multirow{3}{*}{Approaches} & \multirow{3}{*}{ASR Model} & \multirow{3}{*}{Feature} & \multicolumn{4}{c|}{WER $\downarrow$} & \multirow{3}{*}{\shortstack{Feature\\Bitrate \\ \raisebox{0.3ex}{$\downarrow$}}} \\
  & & & \multicolumn{2}{c|}{MyST} & \multicolumn{2}{c|}{OGI} & \\
 & & & dev & test & dev & test &  \\ 
\midrule
\multirow{3}{*}{Fine-tuned SSL SFMs~\cite{fan24b_interspeech}} & Wav2vec 2.0-CTC 
 & - & 10.6 & 11.1 & 2.1 & 2.5 & - \\
 & HuBERT-CTC & - & 10.5 & 11.3 & 2.2 & 2.5 & - \\
 & WavLM-CTC & - & 9.6 & 10.4 & 1.7 & 1.8 & - \\
\midrule
\multirow{9}{*}{\shortstack{Feature (Continuous/Discrete) \\+ Downstream ASR}} & \multirow{9}{*}{\shortstack{E-Branchformer-CTC}} & \multicolumn{6}{c}{\textit{Continuous Feature \& Representation}} \\
\cline{3-8}
 & & Fbank & 16.4 & 16.5 & 2.2 & 2.9 & 256000 \\ 
 & & WavLM & 10.9 & 11.6 & 3.3 & 3.9 & 1638400 \\
 & & Fine-tuned WavLM & 9.6 & 10.1 & 1.6 & 1.7 & 1638400 \\
\cline{3-8}
 & & \multicolumn{6}{c}{\textit{Unsupervised Semantic Tokens}} \\
\cline{3-8}
 & & WavLM K-means & 11.4 & 11.9 & 5.0 & 6.4 & \textbf{548.3} \\
 & & Fine-tuned WavLM K-means & 10.1 & 10.7 & 2.8 & 3.1 & \textbf{548.3} \\
\cline{3-8}
 & & \multicolumn{6}{c}{\textit{{Supervised Semantic Tokens}}} \\
\cline{3-8}
 & & Fine-tuned WavLM FSQ & \textbf{9.3}$^*$ & \textbf{10.0}$^*$ & \textbf{1.5} & \textbf{1.5}$^*$ & \textbf{548.3} \\
\bottomrule
\end{tabular}}
\begin{tablenotes}
    \footnotesize
		\item $^*$: Statistical significance is confirmed with $p < 0.05$
\end{tablenotes}
\label{tab:overall}
\end{table*}

\vspace{0.2cm}
\section{Experimental Setups}
\vspace{0.1cm}
\subsection{Dataset}
\vspace{0.1cm}


To ensure a fair and consistent comparison with previous benchmark studies, we conduct experiments on two child speech corpora as used in~\cite{fan24b_interspeech}: the My Science Tutor (MyST) corpus of \textbf{spontaneous} speech~\cite{WardCBBSVWZB11}, and the CSLU OGI Kids (OGI) corpus of \textbf{scripted}, read speech~\cite{ShobakiHC00}. We follow the preprocessing procedure described in~\cite{fan24b_interspeech}.

The MyST corpus comprises 240 hours of speech transcribed from children in grades 3–5 (aged 8–10 years) recorded during virtual tutoring sessions. Following~\cite{AttiaL0DE24}, we perform quality filtering using Whisper-largeV2, and discard utterances that yield a WER greater than 50\% or contain fewer than three words. We also exclude utterances longer than 30 seconds. After filtering, we obtain 133 hours of training data, with development and test sets comprising 21 and 25 hours, respectively.

The OGI Kids corpus comprises 50 hours of speech collected from 1,100 children aged 4–15, reading isolated words, sentences, or sequences of digits. The data is divided into training (70\%), development (15\%), and test (15\%) sets, with no speaker overlap across splits~\cite{fan24b_interspeech}.


\vspace{0.2cm}
\subsection{Configurations}
\vspace{0.2cm}


\footnotetext[1]{\url{https://huggingface.co/FSQChildASR/wavlm-large-myst}}
\footnotetext[2]{\url{https://huggingface.co/FSQChildASR/wavlm-large-ogi}}

For the speech foundation model, we adopt WavLM Large~\cite{ChenWCWLCLKYXWZ22}, a leading self-supervised learning (SSL) model, as the backbone for both the unsupervised K-means tokenizer and the supervised FSQ tokenizer. K-means clustering is performed with the default setting of 2000 clusters, following~\cite{chang24b_interspeech}. For FSQ, we configure the codebook granularity as $\mathcal{L} = \left[5, 5, 5, 4, 4\right]$, yielding a total codebook size of 2000, thereby aligning with the number of K-means clusters for fair comparison.

Both K-means and FSQ tokenizers are trained on the full training set of each dataset. For K-means, we extract speech representations from the 21st layer of a pre-trained WavLM model and the final (24th) layer of a WavLM model fine-tuned on the respective training set$^{1,2}$. For FSQ training, we initialize the model with a fine-tuned WavLM and optimize it using CTC loss~\cite{GravesFGS06}, during which the transformer layers in WavLM are jointly trained.

We use ASR as the downstream evaluation task throughout this work. For consistency, we adopt the default 12-layer E-Branchformer architecture~\cite{KimWPPSHW22} from ESPnet~\cite{WatanabeHKHNUSH18}, and perform CTC-only decoding using greedy search. When using discrete speech tokens as input, their embeddings are randomly initialized. For text targets, we employ byte pair encoding (BPE) with a vocabulary size of 5000 on MyST, and a smaller size of 200 for OGI to reflect its limited lexical diversity.

\vspace{0.2cm}
\subsection{Metrics}
\vspace{0.1cm}

In addition to Word Error Rate (WER) as the primary evaluation metric, we also report bitrate (bits per second) to provide an auxiliary measure of the encoding efficiency of different continuous and discrete speech representations. The bitrate is computed following the methodology outlined in~\cite{chang24b_interspeech}, allowing a consistent and fair comparison across various encoding schemes.

\vspace{0.2cm}
\section{Experimental Results}
\vspace{0.2cm}
\subsection{Overall Comparison on Child ASR}
\vspace{0.2cm}

Table~\ref{tab:overall} compares various approaches on the MyST and OGI datasets. Baseline results from prior work, using fine-tuned self-supervised learning (SSL) speech foundation models (SFMs) for direct inference, are reported in the first three rows. The remaining rows evaluate different input representations, both continuous and discrete, within the downstream E-Branchformer-CTC ASR model.

Among these approaches, Fbank does not show promising results due to the lack of pre-training benefits. In contrast, using WavLM representations, especially those fine-tuned on in-domain data, yields substantial performance improvements, achieving WERs of 9.6\% / 10.1\% on MyST-dev/test and 1.6\% / 1.7\% on OGI-dev/test. However, the high dimensionality of WavLM Large (1024) combined with its 32-bit floating-point representation leads to an exceptionally high bitrate of 1,638,400 bits per second.

Discrete semantic tokens offer a compelling alternative by combining the benefits of SFM pre-training with efficient compression. Notably, supervised FSQ tokens achieve the best overall performance, reaching WERs of 9.3\% / 10.0\% on MyST and 1.5\% / 1.5\% on OGI, outperforming both Fbank and unsupervised K-means tokens. Remarkably, FSQ tokens  \textbf{outperform in-domain fine-tuned WavLM representations} as well as the direct inference results from fine-tuned SSL SFMs, despite operating at a much lower bitrate and using the same WavLM backbone.

{This finding is counterintuitive, as continuous upstream features are typically expected to outperform their discrete counterparts in ASR due to information loss during discretization~\cite{ChangYFM023,ChangYCJLMSST0F24,YangSDM0P024,chang24b_interspeech}.} We hypothesize that this advantage arises because the WavLM backbone used in supervised FSQ tokenizer was explicitly adapted during supervised training to facilitate discretization, thereby enhancing its effectiveness for downstream ASR tasks.

\subsection{Comparison of De-duplication and Sub-word Modeling on Discrete Speech Tokens}

Table~\ref{tab:strategy} presents the ASR performance and bitrate when applying de-duplication (DD) and {Byte-Pair Encoding (BPE) sub-word (SW) modeling~\cite{ChangYFM023}} as post-processing strategies on speech token sequences. Both techniques aim to reduce redundancy: DD eliminates consecutive repeated tokens, while SW compresses frequent token patterns into subword units.

The results show that both DD and SW modeling effectively reduce the bitrate by shortening the overall token sequence length. However, this compression-induced reduction leads to a slight degradation in ASR performance, highlighting a trade-off between compactness and recognition accuracy. Notably, applying DD to supervised FSQ tokens incurs only a modest accuracy loss while achieving a meaningful reduction in bitrate.

Across all post-processing configurations, FSQ tokens consistently achieve a lower bitrate than K-means tokens, indicating greater efficiency in discarding redundant information during compression. These findings suggest that supervised FSQ tokens provide superior encoding efficiency by preserving essential semantic content while minimizing redundancy.

\vspace{0.2cm}
\subsection{Frequency Distribution Analysis of Speech Tokens}
\label{sec:distribution}
\begin{figure}[t]
    \centering
    \captionsetup[subfigure]{justification=centering} 
    \subfloat[Distribution of unsupervised K-means tokens on MyST]{
        \includegraphics[width=0.46\textwidth]{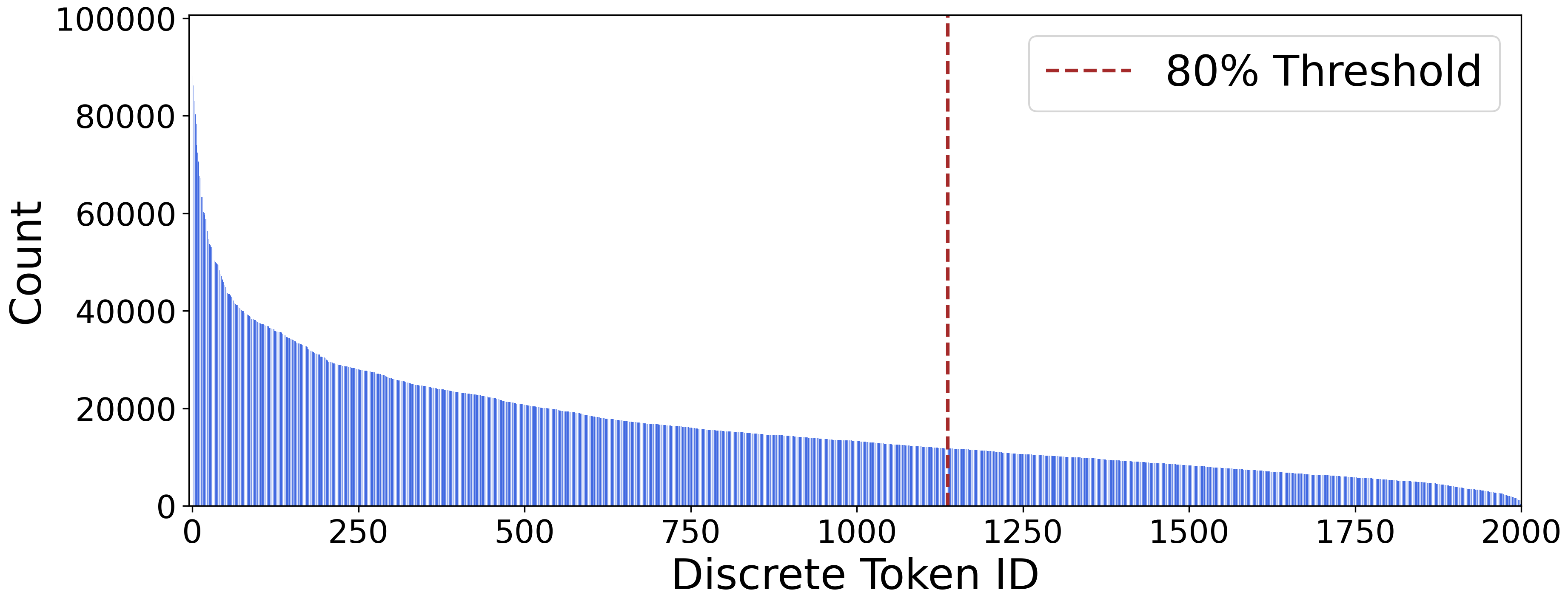}
    }
    
    \vskip 1em
    
    \subfloat[Distribution of supervised FSQ tokens on MyST]{
        \includegraphics[width=0.45\textwidth]{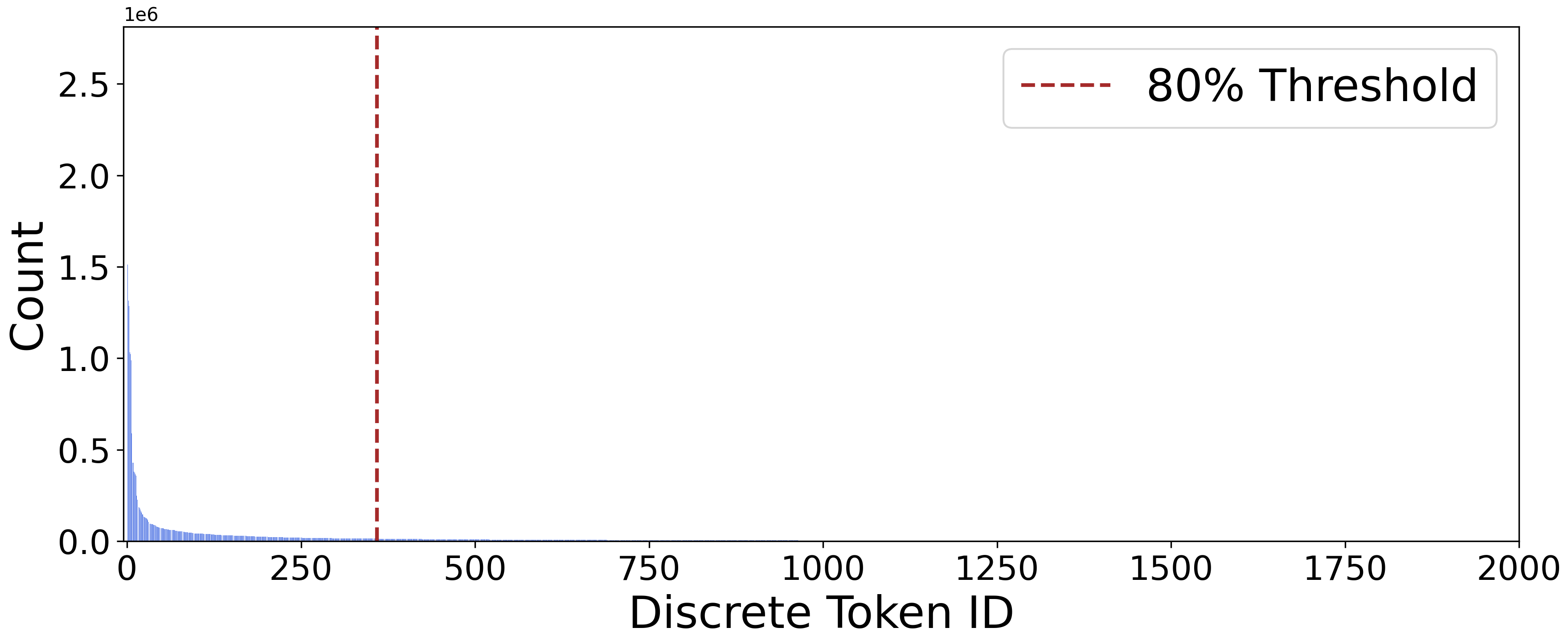}
    }
    \caption{Frequency distribution of discrete tokens with a codebook size of 2000, based on the MyST corpus. The y-axis indicates the frequency of tokens, and the x-axis denotes token IDs sorted in descending order of frequency. The red line marks the 80\% cumulative frequency threshold.}
    \label{fig:distribution}
\end{figure}

\vspace{0.2cm}
Figure~\ref{fig:distribution} illustrates the frequency distribution of both types of discrete tokens on the MyST dataset. Interestingly, we observe that K-means tokens are relatively uniformly distributed across the codebook, whereas FSQ tokens exhibit a sharp and highly skewed distribution, with a large proportion of occurrences concentrated in a small subset of tokens.

\begin{table}[t]
\centering
\caption{Comparison of WER (\%) and Bitrate (bits/sec) using de-duplication (DD) and BPE sub-word (SW) modeling on speech tokens. Bitrate with DD and SW modeling is based on MyST, with an SW vocabulary of 6000 for a codebook size of 2000.}
{
\begin{tabular}{c|c|c|cc|cc|c}
\toprule
\multirow{3}{*}{\shortstack{Token\\Type}} & \multirow{3}{*}{DD} & \multirow{3}{*}{SW} & \multicolumn{4}{c|}{WER $\downarrow$} & \multirow{3}{*}{Bitrate $\downarrow$} \\
 & & & \multicolumn{2}{c|}{MyST} & \multicolumn{2}{c|}{OGI} & \\
& & & dev & test & dev & test &  \\ 
\midrule
\multirow{4}{*}{{K-means}} & \xmark & \xmark & 10.1 & 10.7 & 2.8 & 3.1 & 548.3 \\
 & \cmark & \xmark & 10.3 & 10.8 & 2.9 & 3.0 & 388.0 \\
 & \xmark & \cmark & 10.4 & 10.8 & 3.0 & 3.1 & 351.2 \\
 & \cmark & \cmark & 10.4 & 10.9 & 6.2 & 6.1 & 267.6 \\
 \midrule
\multirow{4}{*}{{FSQ}} & \xmark & \xmark & \textbf{9.3}$^*$ & \textbf{10.0}$^*$ & \textbf{1.5} & \textbf{1.5}$^*$ & 548.3 \\
 & \cmark & \xmark & 9.4 & 10.2 & \textbf{1.5} & 1.7 & 329.1 \\
 & \xmark & \cmark & 9.4 & 10.1 & 4.6 & 4.5 & 304.3 \\
 & \cmark & \cmark & 9.5 & 10.2 & 5.6 & 5.1 & \textbf{239.2} \\
\bottomrule
\end{tabular}}
\begin{tablenotes}
    \footnotesize
		\item $^*$: Statistical significance is confirmed with $p < 0.05$
\end{tablenotes}
\label{tab:strategy}
\vspace{0.5cm}
\end{table}

Remarkably, this sharp distribution correlates with better ASR performance, {challenging the commonly held assumption that a more uniform token usage is beneficial~\cite{abs-2411-08742,abs-2412-10117}.} We hypothesize that this phenomenon may be attributed to the nature of speech phonemes, where only a limited set of sound categories including vowels, consonants and semi-vowels need to be effectively represented for successful transcription. As a result, a non-uniform token distribution that prioritizes frequently occurring phonetic units might actually enhance ASR performance.
This observation motivates our further investigation into whether reducing the codebook size of discrete tokens can maintain or even improve ASR performance, as discussed in Section~\ref{sec:codesize}.

\subsection{Exploring Ultra-Low Bitrate Speech Tokens}
\label{sec:codesize}


Building on the highly skewed FSQ token distribution observed in Section~\ref{sec:distribution}, we explore the impact of reducing the codebook size to a significantly lower level. Specifically, we set $\mathcal{L} = [8,6,5]$, resulting in a codebook of size 240 (approximately $2^8$), which is also the default configuration in~\cite{MentzerMAT24}. For fair comparison, we configure K-means clustering to produce 240 centers as well. The results are presented in Table~\ref{tab:codesize}, with the first row showing continuous fine-tuned WavLM representation as a reference baseline.

Surprisingly, both tokenizers with a codebook of only 240 entries achieve ASR performance comparable to that of the original 2000-entry codebook across both datasets. For K-means tokens, whose usage is more uniformly distributed, reducing the codebook size to 240 leads to a slight drop in ASR accuracy on the MyST test set, whereas it yields an improvement on the OGI dataset. FSQ tokens also achieve comparable or better performance on OGI. We attribute this to the characteristics of the OGI corpus, which involves younger child speakers with more constrained vocabularies, a smaller dataset size, and simpler scripted utterances. These factors inherently require fewer distinct token types, making a reduced codebook size more effective by eliminating redundant tokens.

These results reinforce our earlier observation (Section~\ref{sec:distribution}) that only a compact subset of semantic speech units may be sufficient for effective ASR. Moreover, applying additional techniques such as token de-duplication (DD) and BPE-based sub-word (SW) modeling enables further bitrate reduction with minimal impact on transcription quality.

\begin{table}[t]
\centering
\caption{Comparison of WER (\%) and bitrate (bits/sec) for codebook sizes of 2000 and 240 on the MyST and OGI datasets. Bitrate with de-duplication (DD) and BPE sub-word (SW) modeling is reported on MyST, with SW vocabularies of 6000 (for 2000) and 600 (for 240), respectively.}
\resizebox{0.9\columnwidth}{!}{
\begin{tabular}{l|cc|cc|c}
\toprule
\multirow{3}{*}{Codebook Size} &  \multicolumn{4}{c|}{WER $\downarrow$} & \multirow{3}{*}{Bitrate $\downarrow$} \\
 & \multicolumn{2}{c|}{MyST\footnotemark[1]} & \multicolumn{2}{c|}{OGI\footnotemark[2]} & \\
& dev & test & dev & test & \\
\midrule
\multicolumn{6}{c}{\textit{Continuous Fine-tuned WavLM Representation}} \\
\midrule
\multicolumn{1}{c|}{-} & 9.6 & 10.1 & 1.6 & 1.7 & 1638400 \\
\midrule
\multicolumn{6}{c}{\textit{Unsupervised K-means Tokens}} \\
\midrule
2000 & 10.1 & 10.7 & 2.8 & 3.1 & 548.3 \\
240 & 10.3 & 10.9 & 1.9 & 2.2 & 395.3 \\
\midrule
\multicolumn{6}{c}{\textit{Supervised FSQ Tokens}} \\
\midrule
2000 & \textbf{9.3} & \textbf{10.0}$^*$ & 1.5 & \textbf{1.5} & 548.3 \\
240 & 9.4 & 10.2 & \textbf{1.4} & \textbf{1.5} & 395.3 \\
\enspace + DD & 9.5 & 10.2 & 1.6 & 1.8 & 224.2 \\
\quad + SW & 9.6 & 10.3 & 2.0 & 2.2 & \textbf{186.9} \\
\bottomrule
\end{tabular}}
\begin{tablenotes}
    \footnotesize
		\item $^*$: Statistical significance is confirmed with $p < 0.05$
\end{tablenotes}
\label{tab:codesize}
\end{table}

\subsection{Cross-Domain Evaluation of the Speech Tokenizers}

We further assess the robustness and generalizability of speech tokenizers across different age groups and styles using different domains of tokenizers. Specifically, we define the domain of a speech tokenizer based on its training data (e.g., ``Pretrained SSL'' means that tokens are extracted from a pretrained SSL model). We train on the combined MyST and OGI training sets using each domain of speech tokenizers, and evaluate on test sets across different styles (the scripted OGI test set and the spontaneous MyST test set) and age groups (4–7, 8–10, and 11–15 years old). The WER results are summarized in Table~\ref{tab:cross}.

As expected, the best performance on both test set styles is achieved using an in-domain tokenizer, while noticeable degradation is observed when using out-of-domain tokenizers. Notably, using a pretrained SSL model to extract speech tokens maintains a relatively good balance across the two test set styles, as it is not fine-tuned for any specific domain. Using tokenizers fine-tuned on spontaneous child speech can improve performance across both styles while still maintaining a good balance. In contrast, tokenizers fine-tuned on scripted child speech improve performance on the OGI scripted test set but significantly degrade performance on the spontaneous test set. This is because these tokenizers are overfitted to scripted child speech data, which features a more limited and less diverse vocabulary and generalizes poorly to other data.


Furthermore, while the supervised FSQ tokenizer yields overall better ASR results in the in-domain setting, its performance suffers more significantly in cross-domain scenarios compared to the unsupervised K-means tokenizer. This suggests that the supervised FSQ tokenizer may be more closely aligned with the specific characteristics of the training data, which in turn reduces its generalizability to other domains. In contrast, the K-means tokenizer is trained solely based on the distance relationships between frame-level representations, without incorporating any task-specific objectives.

Finally, for different age groups, we observe that younger children’s speech (4–7 years) is more challenging than that of older children. Due to the imperfect development of the vocal tract, their pronunciation differs significantly from older age groups. These findings highlight the importance of developing discrete speech tokenizers that are not only effective within a domain but also robust and adaptable across diverse speech styles and age groups. Future research may explore domain-agnostic and age-invariant training strategies and representations to enhance transferability in practical applications.

\begin{table}[t]
\centering
\caption{Evaluation of speech tokenizers trained on data from different domains, evaluated across styles (Scripted, Spontaneous) and age groups (4–7, 8–10, 11–15 years) in terms of WER (\%).}
\resizebox{\columnwidth}{!}{
\begin{tabular}{c|ccc|c}
\toprule
\multirow{3}{*}{} &  \multicolumn{4}{c}{WER $\downarrow$} \\
 & \multicolumn{3}{c|}{OGI (Scripted)} & MyST (Spon) \\
& Age 4-7 & Age 8-10 & Age 11-15 & Age 8-10 \\
\midrule
\multicolumn{5}{l}{\textit{Training Domain of Unsupervised K-means Tokens}} \\
\midrule
Pretrained SSL & 11.0 & 2.5 & 1.8 & 11.7 \\
MyST (Spon) & 10.1 & 1.9 & 1.3 & 10.7 \\
OGI (Scripted) & 7.3 & 1.0 & 0.9 & 78.2 \\
\midrule
\multicolumn{5}{l}{\textit{Training Domain of Supervised FSQ Tokens}} \\
\midrule
MyST (Spon) & 12.0 & 2.4 & 1.4 & \textbf{10.0}$^*$ \\
OGI (Scripted) & \textbf{6.3}$^*$ & \textbf{0.8}$^*$ & \textbf{0.6}$^*$ & 88.0 \\
\bottomrule
\end{tabular}}
\begin{tablenotes}
    \footnotesize
		\item $^*$: Statistical significance is confirmed with $p < 0.05$
\end{tablenotes}
\label{tab:cross}
\end{table}

\vspace{0.3cm}
\section{Conclusion}
\vspace{0.2cm}
{This paper presents a comprehensive comparative analysis between K-means-based unsupervised and FSQ-based supervised semantic speech tokens for low-resource child ASR, evaluating their ASR performance, bitrate efficiency, codebook distribution, and cross-domain generalization.} Experiments on the MyST and OGI datasets demonstrate that FSQ-based supervised speech tokens not only surpass K-means-based unsupervised tokens but also unexpectedly outperform continuous representations from SFMs. Experimental results and analysis of codebook distributions further reveal that discrete speech tokens remain effective on child ASR even in ultra-low bitrate settings. Cross-domain evaluations, however, highlight a key limitation: both tokenizers show limited generalizability across speaking styles and age groups. In future work, we plan to investigate alternative speech foundation models and datasets to obtain more effective semantic tokens and to develop unified, robust, and efficient speech tokenizers aimed at improving both performance and generalization.

\newpage


\end{document}